\documentclass[12pt]{article}
\newcommand{\nn}{\nonumber\\}

\newcommand{\C}{{\bf C}}

\newcommand{\del}{\partial}
\newcommand{\vvert}{\Big{\vert}}

\newcommand{\tr}{{\rm tr}\,}

\newcommand{\rar}{\rightarrow}

\newcommand{\fr}{\frac}

\newcommand{\zb}{\bar{z}}

\newcommand{\res}{\mbox{res}}
\makeatletter
\@addtoreset{equation}{section}

\makeatother

\usepackage{graphicx}

\begin{document}

\begin{titlepage}
\null
\begin{flushright}
arXiv:yymm.nnnn
\\
December, 2010
\end{flushright}

\vskip 1.5cm
\begin{center}

 {\Large \bf Noncommutative Integrable Systems and Quasideterminants}

\vskip 1.7cm
\normalsize

{\large Masashi Hamanaka\footnote{
E-mail: hamanaka@math.nagoya-u.ac.jp}}

\vskip 1.5cm

        {\it Graduate School of Mathematics, Nagoya University,\\
                     Chikusa-ku, Nagoya, 464-8602, JAPAN}

\vskip 1.5cm

{\bf \large Abstract}

\end{center}

We discuss extension of soliton theories and
integrable systems into noncommutative spaces.
In the framework of noncommutative integrable hierarchy,
we give infinite conserved quantities and exact soliton solutions
for many noncommutative integrable equations, which are represented 
in terms of Strachan's products and quasi-determinants, respectively.
We also present a relation to an noncommutative anti-self-dual Yang-Mills
equation, and make comments on how ``integrability'' 
should be considered in noncommutative spaces.


\end{titlepage}
 
\section{Introduction}

Noncommutative extension of field theories is not just
a generalization of them but a fruitful study direction
in both physics and mathematics. First of all, 
we introduce motivation and goal of it.

\subsection{Motivation to extend to noncommutative spaces}

Noncommutative spaces are characterized 
by the noncommutativity of the spatial coordinates $x^\mu$:
\begin{eqnarray}
\label{nc_coord}
[x^\mu,x^\nu]=i\theta^{\mu\nu},
\end{eqnarray}
where the anti-symmetric tensor $\theta^{\mu\nu}$ is 
called the {\it noncommutative parameter}.
In this article, the noncommutative parameter is a real constant and  
closely related to existence of a background flux.

We summarize some properties of field theories on noncommutative spaces.

\begin{itemize}

 \item Resolution of singularities

Eq. (\ref{nc_coord}) looks like the canonical commutation
relation $[q,p]=i\hbar$ in quantum mechanics
and would lead to ``space-space uncertainty relation.''
Hence the singularity which exists on commutative spaces
could resolve on noncommutative spaces.
This is one of the distinguished features of noncommutative theories
and gives rise to various new physical objects
such as $U(1)$ instantons \cite{NeSc}.

\item Equivalence between noncommutative gauge theory and commutative gauge
theory in background magnetic fields

In the context of effective theory of D-branes,  
noncommutative gauge theories are found to be equivalent to ordinary 
gauge theories in the presence of background magnetic fields
and have been studied intensively for
the last several years (For reviews, see e.g. \cite{DoNe,KoSc,Szabo}.)
Noncommutative solitons especially play important roles in the study of 
D-brane dynamics, such as the
confirmation of Sen's conjecture on tachyon condensation. (For
reviews, see e.g. \cite{Hamanaka3, Harvey2,Sen}.) 
We note that $U(1)$ part of the gauge group is necessary and 
plays important roles as in $U(1)$ instantons. 

\item Easy Treatment 

Solitons special to noncommutative spaces are sometimes so simple
that we can calculate various physical quantities,
such as the energy, the fluctuation around
the soliton configuration and so on.
Because of resolution of singularities, singular configurations
becomes smooth and become suitable for the usual calculation.
Furthermore, we can take large noncommutativity limit
where the situations become simple.
The successful application to D-brane dynamics
are actually due to this point. (For a review, see e.g. \cite{Harvey2}.)

\end{itemize}

\subsection{Towards noncommutative integrable systems}

Noncommutative extension of integrable equations such as the KdV equation is
also one of the hot topics. (For reviews, 
see e.g. \cite{DiMH8,Hamanaka5,Lechtenfeld2,Lechtenfeld3,Mazzanti,Tamassia}.) 
These equations imply no gauge field
and noncommutative extension of them perhaps might have no physical picture or
no good property on integrability. To make matters worse, noncommutative
extension of $(1+1)$-dimensional equations introduces infinite
number of time derivatives, which makes it hard to discuss or
define the integrability. Those equations had been examined one by one. 
Hence we proposed the following study programs 
to solve the above problems in more general 
geometrical and physical frameworks:

\begin{itemize}

\item Noncommutative twistor theory together with noncommutative Ward's conjecture

Twistor theory is the most essential framework in
the study of integrability of anti-self-dual Yang-Mills eqs.
(See, e.g. \cite{MaWo,WaWe}.) 
Noncommutative extension of twistor theories are
already discussed by several authors, e.g.
\cite{BrMa,Hannabuss,IhUh,KKO,Takasaki}. This would give a
geometrical foundation of integrability of anti-self-dual Yang-Mills eqs.

The Ward conjecture is a statement that many integrable
equations can be derived from anti-self-dual Yang-Mills eqs. by reduction \cite{Ward}.
Noncommutative Ward's conjecture is very important to give physical pictures to
lower-dimensional integrable equations and to make it possible to
apply analysis of noncommutative solitons to that of the corresponding
D-branes, which is first proposed explicitly by \cite{HaTo}.
(See also \cite{LPS}.)
Origin of the integrable-like properties would be also
revealed from the viewpoints of noncommutative twistor theory.

\item Noncommutative Sato's theory 

Sato's theory is known to be one of the most beautiful theories of
solitons and reveals essential aspects of the integrability, such
as, the construction of exact multi-soliton solutions, the
structure of the solution space, the existence of infinite
conserved quantities, and the hidden symmetry of them, on
commutative spaces. So it is reasonable to extend Sato's theory
onto noncommutative spaces in order to clarify various integrable-like aspects directly.

\end{itemize}

In this article, we report recent developments of noncommutative extension of
soliton theories and integrable systems.
We prove the existence of infinite conserved quantities and 
exact multi-soliton solutions
in the framework of noncommutative integrable hierarchy.
We also give an example of reduction of 
noncommutative Anti-Self-Dual Yang-Mills equation into noncommutative KdV eq.
``Integrability'' in noncommutative spaces is also discussed.

\subsection{Noncommutative Field Equations in the sense of Moyal deformations}

Noncommutative field theories on flat spaces 
are given by the replacement of ordinary products
in the commutative field theories with the {\it Moyal-products} 
\cite{Moyal} and realized as deformed theories from the commutative ones.
The Moyal-product is defined for ordinary fields explicitly by
\begin{eqnarray}
\label{star}
f\star g(x)&:=&
\exp{\left(\frac{i}{2}\theta^{\mu\nu}\partial^{(x^\prime)}_\mu
\partial^{(x^{\prime\prime})}_\nu\right)}
f(x^\prime)g(x^{\prime\prime})\Big{\vert}_{x^{\prime}
=x^{\prime\prime}=x}\nonumber\\
&=&f(x)g(x)+\frac{i}{2}\theta^{\mu\nu}\partial_\mu f(x)\partial_\nu g(x)
+{\cal O}(\theta^2),
\end{eqnarray}
where $\del_i^{(x^\prime)}:=\del/\del x^{\prime i}$
and so on. 
The Moyal-product has associativity: $f\star(g\star h)=(f\star g)\star h$,
and reduces to the ordinary product in the commutative limit:
 $\theta^{\mu\nu}\rar 0$.
The modification of the product  makes the ordinary
spatial coordinate ``noncommutative'' which means :
$[x^\mu,x^\nu]_\star:=x^\mu\star x^\nu-x^\nu\star x^\mu=i\theta^{\mu\nu}$.

We note that the fields themselves take c-number values
as usual and the differentiation and the integration for them
are well-defined as usual. 
A nontrivial point is that
noncommutative field equations contain infinite number
of derivatives in general.
Hence the integrability of the equations
are not so trivial as commutative cases,
especially for space-time noncommutativity.

In this article, we mainly studies noncommutative KP and KdV equations:
\begin{itemize}

 \item Noncommutative KP equation in $(2+1)$-dimension ($[x,y]_\star=i\theta$
or $[t,x]_\star=i\theta$)

\begin{eqnarray}
 \fr{\del u}{\del t}=\frac{1}{4}\fr{\del^3 u}{\del x^3}
+\frac{3}{4}\left(\fr{\del u}{\del x}\star u+u\star \fr{\del u}{\del x}\right)
+\frac{3}{4}\partial_x^{-1} \fr{\del^2 u}{\del y^2}
-\frac{3}{4}\left[u,\partial_x^{-1} \fr{\del u}{\del
y}\right]_\star,
\label{kp}
\end{eqnarray}
where $t$ and $x,y$ are time and spatial coordinates, respectively,
and $\partial_x^{-1}f(x)=\int^x dx^\prime f(x^\prime)$.

\item Noncommutative KdV equation in $(1+1)$-dimension ($[t,x]_\star=i\theta$)

\begin{eqnarray}
 \fr{\del u}{\del t}=\frac{1}{4}\fr{\del^3 u}{\del x^3}
+\frac{3}{4}\left(\fr{\del u}{\del x}\star u+u\star \fr{\del u}{\del x}\right).
\label{kdv}
\end{eqnarray}

\end{itemize}

The ordering of non-linear terms is crucial to preserve
some special integrable properties and determined in the Lax
formalism as we will see later. For noncommutative KP and KdV eqs.,
the non-linear term $2u\cdot \del_xu$ becomes symmetric: 
$\del_xu\star u+u\star \del_x u $. 

\section{Noncommutative integrable systems}

In this section, we discuss some integrable aspects
of noncommutative integrable equations focusing on noncommutative KdV eq.


\subsection{Noncommutative integrable Hierarchies}

Firstly, we derive various noncommutative integrable equations 
in terms of pseudo-differential operators 
which include negative powers of differential operators.

An $n$-th order pseudo-differential operator $A$
is represented as follows
\begin{eqnarray}
\label{pseudo}
 A=a_n \del_x^n + a_{n-1}\del_x^{n-1}+ \cdots
+ a_0 +a_{-1}\del_x^{-1}+a_{-2}\del_x^{-2}+\cdots,
\end{eqnarray}
where $a_i$ is a function of $x$ 
associated with noncommutative associative products
(here, the Moyal products).
When the coefficient of the highest order $a_n$ equals to 1,
we call it {\it monic}.
Here we introduce useful symbols:
\begin{eqnarray}
 A_{\geq r}&:=& \del_x^n + a_{n-1}\del_x^{n-1}+ \cdots + a_{r}\del_x^{r},\\
 A_{\leq r}&:=& A - A_{\geq r+1}
 = a_{r}\del_x^{r} + a_{r-1}\del_x^{r-1} +\cdots.\\
 \res_{r} A &:=& a_{r}.
\end{eqnarray}
The symbol $\res_{-1} A$ is especially called the {\it residue} of $A$.

The action of a differential operator $\partial_x^n$ on
a multiplicity operator $f$ is formally defined
as the following generalized Leibniz rule:
\begin{eqnarray}
 \partial_x^{n}\cdot f:=\sum_{i\geq 0}
\left(\begin{array}{c}n\\i\end{array}\right)
(\partial_x^i f)\partial^{n-i},
\end{eqnarray}
where the binomial coefficient is given by
\begin{eqnarray}
\label{binomial}
 \left(\begin{array}{c}n\\i\end{array}\right):=
\frac{n(n-1)\cdots (n-i+1)}{i(i-1)\cdots 1}.
\end{eqnarray}
We note that the definition of the binomial coefficient (\ref{binomial})
is applicable to the case for negative $n$,
which just define the action of
negative power of differential operators.

The composition of pseudo-differential operators
is also well-defined and the total set
of pseudo-differential operators forms
an operator algebra.
For more on pseudo-differential operators
and integrable hierarchies, 
see e.g. \cite{BBT, Blaszak, Dickey, Kupershmidt}.

\vspace{3mm}

In order to define the noncommutative KP hierarchy,
let us introduce a Lax operator:
\begin{eqnarray}
 L = \partial_x + u_2 \partial_x^{-1}
 + u_3 \partial_x^{-2} + u_4 \partial_x^{-3} + \cdots,
~~~u_k=u_k(x; x^1,x^2,x^3, \ldots).
\end{eqnarray}
The noncommutativity is introduced into
the coordinates $(x^1,x^2,\ldots)$.
The differential operator $B_m$ is given by
\begin{eqnarray}
 B_m:=(\underbrace{L\star \cdots \star L}_{ m{\scriptsize\mbox{
     times}}})_{\geq 0}.
\end{eqnarray}

The noncommutative KP hierarchy is defined as
\begin{eqnarray}
 \del_m L = \left[B_m, L\right]_\star,~~~m=1,2,\ldots,
\label{lax_sato}
\end{eqnarray}
where the action of $\del_m:=\del/\del x^m$
on the pseudo-differential operator $L$
should be interpreted to be coefficient-wise,
that is, $\del_m L :=[\del_m,L]_\star$ or $\del_m \del_x^k=0$.
The KP hierarchy gives rise to a set of infinite differential
equations with respect to infinite kind of fields from the
coefficients in Eq. (\ref{lax_sato}) for a fixed $m$. Hence it
contains huge amount of differential (evolution) equations for all
$m$. The LHS of Eq. (\ref{lax_sato}) becomes $\del_m u_k$ which
shows a kind of flow in the $x_m$ direction. 

If we put the constraint $(L^l)_{\leq -1}=0$ or equivalently $L^l=B_l$ 
on the noncommutative KP hierarchy (\ref{lax_sato}), 
we get a reduced noncommutative KP hierarchy which is called 
the {\it l-reduction} of the noncommutative KP hierarchy, or
the {\it noncommutative $l$KdV hierarchy}, or the $l$-th 
{\it noncommutative Gelfand-Dickey hierarchy}. 
We can easily show
$\partial u_k / \partial x^{Nl}=0$
for all $N,k$ because
${\partial L^l}/{\partial x^{Nl}}=[B_{Nl},L^l]_\star
=[(L^{l})^N,L^l]_\star=0$. 
In particular, the 2-reduction of the noncommutative KP hierarchy 
is just the noncommutative KdV hierarchy.

{}Let us see explicit examples.
\begin{itemize}

\item Noncommutative KP hierarchy

The coefficients of each powers of (pseudo-)differential operators
in the noncommutative KP hierarchy (\ref{lax_sato}) yield a series of infinite
noncommutative ``evolution equations'' which commute each other. 
For example,
\begin{itemize}
 \item  for $m=1$
\begin{eqnarray}
\partial_x^{1-k})~~~ \del _1 u_{k}=\partial_x u_{k},~~~k=2,3,\ldots.
\end{eqnarray}
Hence we can identify $x^1\equiv x$.
\item for $m=2$
\begin{eqnarray}
\label{KP_hie}
\partial_x^{-1})~~~\del_2 u_{2}
&=&u_2^{\prime\prime}+2u_{3}^{\prime},\nonumber \\
\partial_x^{-2})~~~
\del_2 u_{3}&=&u_3^{\prime\prime}+2u_4^{\prime}
+2u_2\star u_2^\prime +2[u_2,u_3]_\star,\nonumber \\
\partial_x^{-3})~~~
\del_2 u_{4}&=&u_{4}^{\prime\prime}+2u_{5}^{\prime}
+4u_3\star u_2^\prime-2u_2\star u_2^{\prime\prime}
+2[u_2,u_4]_\star,\nn
\partial_x^{-4})~~~\del_2 u_{5}&=&\cdots,
\end{eqnarray}
which implies that infinite kind of fields $u_3, u_4, u_5,\ldots$
are represented in terms of one kind of field  $2u_2\equiv u$
\cite{HaTo3}.
\item for $m=3$
\begin{eqnarray}
\label{3flow}
\partial_x^{-1})~~~
\del_3 u_{2}&=&u_{2}^{\prime\prime\prime}+3u_3^{\prime\prime}
+3u_4^{\prime}+3u_2^\prime\star u_2+3u_2\star u_2^\prime,
\nonumber\\
\partial_x^{-2})~~~
\del_3 u_{3}&=&u_{3}^{\prime\prime\prime}+3u_{4}^{\prime\prime}
+3u_{5}^\prime+6u_{2}\star u_{3}^\prime+3u_2^\prime\star u_3
+3u_3\star u_2^\prime+3[u_2, u_4]_\star,\nn
\partial_x^{-3})~~~
\del_3 u_{4}&=&u_{4}^{\prime\prime\prime}+3u_{5}^{\prime\prime}
+3u_{6}^\prime+3u_{2}^\prime \star u_{4}+3u_2\star u_4^\prime
+6u_4\star u_2^\prime\nn
&&-3u_2\star u_3^{\prime\prime}
-3u_3\star u_2^{\prime\prime}+6u_3\star u_3^{\prime}
+3[u_2,u_5]_\star+3[u_3,u_4]_\star,\nn
\partial_x^{-4})~~~\del_3 u_{5}&=&\cdots.
\end{eqnarray}
These just imply the $(2+1)$-dimensional noncommutative KP equation (\ref{kp})
with $2u_2\equiv u, x^2\equiv
y,x^3\equiv t$.
\end{itemize}

Higher-order flows give an infinite set of 
higher-order noncommutative KP equations. The order of nonlinear 
terms are determined in this way.

\item Noncommutative KdV Hierarchy (2-reduction of the noncommutative KP hierarchy)

Taking the constraint $L^2=B_2=:\del_x^2+u$ for
the noncommutative KP hierarchy, we get the noncommutative KdV hierarchy.
This time, the following noncommutative hierarchy
\begin{eqnarray}
\label{KdV_hie}
 \frac{\partial u}{\partial x^m}=\left[B_m, L^2\right]_\star,
\end{eqnarray}
include neither positive nor negative power of
(pseudo-)differential operators for the same reason as commutative
case and gives rise to the $m$-th KdV equation for each $m$. 
For example, 
\begin{itemize}
 \item for $m=3$, identifying the time coordinate as $x^3\equiv t$:
\begin{eqnarray}
\label{ncKdV}
 \dot{u}\equiv \del_3 u=\frac{1}{4}u^{\prime\prime\prime}+\frac{3}{4}
\left(u^\prime \star u + u \star u^\prime \right),
\end{eqnarray}
which is just the $(1+1)$-dimensional noncommutative KdV equation.

 \item for $m=5$ identifying the time coordinate $x^5\equiv t$:
\begin{eqnarray}
\label{5kdv}
\dot{u}\equiv \del_5 u=\frac{1}{16}u^{\prime\prime\prime\prime\prime}
+\frac{5}{16}(u\star
u^{\prime\prime\prime}+u^{\prime\prime\prime}\star u)
+\frac{5}{8}(u^{\prime}\star u^{\prime}+u\star u\star u)^\prime,
\end{eqnarray}
which is the $(1+1)$-dimensional 5-th noncommutative KdV equation.
\end{itemize}
We note that the time coordinate is defined for
each flow equation. This point is important for 
discussion on conserved quantities of noncommutative integrable equations.

\end{itemize}

In this way, we can generate infinite set of the $l$-reduced noncommutative KP
hierarchies. More explicit examples are seen in e.g. \cite{Hamanaka4}.
(See also \cite{OlSo, Wang, WaWa3}.)
The present discussion is also 
applicable to other noncommutative hierarchies, such as, 
the noncommutative Ablowitz-Kaup-Newell-Segur (AKNS) hierarchy \cite{DiMH6},
the noncommutative Toda field hierarchy \cite{Sakakibara}
the noncommutative toroidal KdV hierarchy \cite{Hamanaka8} and so on.

\subsection{Conservation Laws}

Here we prove the existence of infinite conservation laws for the
wide class of noncommutative soliton equations. The existence of infinite
number of conserved quantities would lead to infinite-dimensional
hidden symmetry from Noether's theorem.

First we would like to comment on conservation laws
of noncommutative field equations \cite{HaTo2,Hamanaka4}.
The discussion is basically the same as commutative case
because both the differentiation and the integration
are the same as commutative ones in the Moyal representation.

Let us suppose the conservation law
\begin{eqnarray}
\partial_t \sigma(t,x^i)=\partial_k J^k(t,x^i),
\end{eqnarray}
where $\sigma(t,x^i)$ and $J^k(t,x^i)$ are called
the {\it conserved density} and the {\it associated flux},
respectively.
The conserved quantity is given by spatial integral
of the conserved density:
\begin{eqnarray}
Q(t)=\int_{\scriptsize\mbox{space}}d^Dx \sigma(t,x^i),
\end{eqnarray}
where the integral $\int_{\scriptsize\mbox{space}}dx^D$
is taken for spatial coordinates and
the surface term of the integrand  $J^k(t,x^i)$ 
is supposed to vanish.


Here let us return back to noncommutative hierarchy. In order to discuss the
conservation laws, we have to specify for what equation the
conservation law is. The specified equation possesses its own 
space and time coordinates in the infinite coordinates $x^1,x^2,x^3,\cdots$.
Identifying $t\equiv x^m$, we can get infinite
conserved densities for the noncommutative hierarchies 
as follows ($n=1,2,\ldots$) \cite{Hamanaka4}:
\begin{eqnarray}
\label{cons}
 \sigma_n={\mbox{res}}_{-1} L^n+\theta^{im}\sum_{k=0}^{m-1}\sum_{l=0}^{k}
\left(\begin{array}{c}k\\l\end{array}\right) 
\del_x^{k-l}{\mbox{res}}_{-(l+1)} L^n
\diamond \del_i {\mbox{res}}_{k} L^m,
\label{conservation}
\end{eqnarray}
where the suffices $i$ must run in the space-time directions only.
The product ``$\diamond$'' is 
commutative and non-associative and
and defined by:
\begin{eqnarray}
 f(x)\diamond g(x)
:=\sum_{s=0}^{\infty}
\fr{(-1)^s}{(2s+1)!}\left(\frac{1}{2}\theta^{\mu\nu}
\del_\mu^{(x^\prime)}\del_\nu^{(x^{\prime\prime})}\right)^{2s}
f(x^\prime)g(x^{\prime\prime})\vvert_{x^\prime=x^{\prime\prime}=x},
\end{eqnarray}
which is called {\it Strachan's product} \cite{Strachan}.

We note that the explicit form (\ref{cons}) is
defined for each equation in the noncommutative integrable hierarchy
where space-time coordinates are specified and noncommutativity
is introduced into the specified space-time coordinates only. 
Hence the conserved density is not common in all noncommutative hierarchy 
equations unlike commutative integrable hierarchy.
For example, conserved densities of the (3-th)
noncommutative KdV eq. (\ref{ncKdV}) and the noncommutative 5-th KdV eq. (\ref{5kdv}) 
are different.

We can easily see that deformation terms appear in the second term
of Eq. (\ref{conservation}) in the case of space-time
noncommutativity. On the other hand, in the case of space-space
noncommutativity, the conserved density is given by the residue of
$L^n$ as commutative case.

For examples, explicit representation of the noncommutative KP equation 
is as follows:
\begin{itemize}
 \item space-space noncommutativity $[x,y]_\star=i\theta$:
\begin{eqnarray}
\sigma_n
={\mbox{res}}_{-1} L^n,
\end{eqnarray}
which is essentially the same as commutative one.
In this case, the equation is the first order differential 
equation w.r.t. time and notion of time evolution and Hamiltonian 
structure are well-defined as in commutative situation.
In particular, the trace of a pseudo-differential operator $A$ 
(\ref{pseudo}) should be defined as $\tr A :=\int dxdx^i \res_{-1} A$,
where the integration $\int dx^i$ must be done 
over all spatial directions.

 \item space-time noncommutativity $[t,x]_\star=i\theta$:
\begin{eqnarray}
\sigma_n
={\mbox{res}}_{-1} L^n
-3\theta
\left(({\mbox{res}}_{-1}L^n)\diamond u_3^\prime
+({\mbox{res}}_{-2}L^n)\diamond u_2^\prime
\right).
\end{eqnarray}
This time, the deformation part is proved to be non-trivial. 
The meaning of the existence of infinite conserved 
quantities is however not yet clarified because the equation
contains infinite time derivatives. It is originally hard to
discuss notion of time evolution, Hamiltonian structure
and so on. One solution is to find
the corresponding commutative description via the 
Seiberg-Witten map \cite{SeWi} for effective gauge theories 
of D-branes.

\end{itemize}

\subsection{Exact Soliton Solutions}

Here we show the existence of exact multi-soliton solutions
of noncommutative integrable hierarchy 
by giving the explicit formula in terms of quasideterminants.

Let us introduce the following functions,
\begin{eqnarray}
 \label{argument}
 f_s(\vec{x})=e_\star^{\xi(\vec{x};\alpha_s)}
  +a_s e_\star^{\xi(\vec{x};\beta_s)},~~~
 \xi(\vec{x};\alpha)=x^1\alpha+x^2 \alpha^2+x^3 \alpha^3+\cdots,
\end{eqnarray}
where $\alpha_s$, $\beta_s$ and $a_s$ are constants. 
Moyal exponential functions are defined by
\begin{eqnarray}
 e_\star^{f(x)}:=1+\sum_{n=1}^{\infty}\frac{1}{n!}
\underbrace{f(x)\star \cdots \star f(x)}_{n ~{\mbox{\scriptsize times}}}.
\end{eqnarray}

An $N$-soliton solution of the noncommutative KP hierarchy (\ref{lax_sato})
is given by a quasideterminant of the Wronski matrix \cite{EGR}:
\begin{eqnarray}
 L=\Phi_N \star \partial_x \Phi_N^{-1},
\label{Nsol1}
\end{eqnarray}
where
\begin{eqnarray}
 \Phi_N \star f&=&\vert W(f_1, \ldots, f_N, f)\vert_{N+1,N+1},\nonumber\\
&=&
\begin{array}{|ccccc|}
 f_1&f_2 & \cdots& f_N & f\\
 f^\prime_1&  f^\prime_2& \cdots& f^\prime_N&f^\prime\\
 \vdots& \vdots&\ddots & \vdots &\vdots\\
 f^{(N-1)}_1& f^{(N-1)}_2& \cdots & f^{(N-1)}_N &f^{(N-1)}\\
 f^{(N)}_1& f^{(N)}_2& \cdots &f^{(N)}_N & \fbox{$f^{(N)}$}\\
      \end{array}~ .
\label{sol_KP}
\end{eqnarray}
Definition of quasideterminants is seen in Appendix A.
The Wronski matrix $W(f_1,f_2,\cdots, f_m)$ is as usual:
\begin{eqnarray}
 W(f_1,f_2,\cdots, f_m):=
\left(\begin{array}{cccc}
 f_1&f_2 & \cdots& f_m\\
 f^\prime_1& f^\prime_2& \cdots& f^\prime_m\\
 \vdots& \vdots&\ddots & \vdots\\
  f^{(m-1)}_1& f^{(m-1)}_2& \cdots & f^{(m-1)}_m\\
      \end{array}\right),
\end{eqnarray}
where $f_1,f_2, \cdots, f_m$ are functions of $x$
and $f^\prime:=\del f/\del x,~
f^{\prime\prime}:=\del^2 f/\del x^2,~
f^{(m)}:=\del^m f/\del x^m$ and so on. 

In the commutative limit, $\Phi_N \star f$ is reduced to
\begin{eqnarray}
 \Phi_N \star f \longrightarrow
  \frac{\det W(f_1,f_2,\ldots,f_N,f)}
{\det W(f_1,f_2,\ldots,f_N)},
\end{eqnarray}
which just coincides with commutative one \cite{Dickey}. 
In this respect, quasi-determinants are fit
to this framework of the Wronskian solutions.

{}From Eq. (\ref{Nsol1}), we have a more explicit form as \cite{EGR,GiNi}:
\begin{eqnarray}
 u_2=\partial_x \left(\sum_{s=1}^{N} W^\prime_s \star W_s^{-1}  \right),
\label{Nsol2}
~~~W_s:=\vert W(f_1,\ldots,f_s)\vert_{ss}.
\end{eqnarray}

The $l$-reduction condition $(L^l)_{\leq -1}=0$ or $L^l=B_l$
is realized at the level of the soliton solutions
by taking $\alpha_s^l=\beta_s^l$ or equivalently
$\alpha_s=\epsilon \beta_s$ for $s=1,\cdots,N$,
where $\epsilon$ is the $l$-th root of unity.
For the KdV eq., $\alpha_s=- \beta_s$.

\bigskip

Physical interpretation of the configurations
is non-trivial because even when $f(x)$ and $g(x)$ are real, 
$f(x)\star g(x)$ is not in general. However, 
the $N$-soliton solutions can be real in the following situations.

\begin{itemize}

 \item One-soliton solutions

First, let us comment on one-soliton solutions \cite{DiMH2,HaTo2}.
Defining $z:=x+vt,
\zb:=x-vt$, we easily see
\begin{eqnarray}
 f(z)\star g(z)= f(z) g(z)
\end{eqnarray}
because the Moyal-product (\ref{star}) is rewritten in terms of
$(z,\zb)$ as
\begin{eqnarray}
 f(z,\zb)\star g(z,\zb)=
e^{iv\theta\left(
\partial_{\zb^\prime}
\partial_{z^{\prime\prime}}-
\partial_{z^\prime}
\partial_{\zb^{\prime\prime}}
\right)}f(z^\prime,\zb^\prime)
g(z^{\prime\prime},\zb^{\prime\prime}) \Big{\vert}_{\scriptsize
z^{\prime}
=z^{\prime\prime}=z,
\zb^{\prime} =\zb^{\prime\prime}=\zb.}
\end{eqnarray}
Hence noncommutative one soliton-solutions are essentially the same as 
commutative ones and hence can be real in all region of the space-time.

 \item Asymptotic region of $N$-soliton solutions

In order to analyze the asymptotic behavior of $N$-soliton solutions, 
we usually take a new coordinate comoving
with the $I$-th soliton. Then we can see that in the asymptotic 
region, the configuration just coincides with the commutative one.
Hence, asymptotic behavior of the multi-soliton solutions 
is all the same as commutative one. As the results,
the $N$-soliton solutions possess $N$ localized energy densities.
In the general scattering process without resonances, they never decay and
preserve their shapes and velocities of the localized solitary waves.
The phase shifts also occur by the same degree as commutative ones.
These observations are crucially due to special properties
of quasideterminants and would be true of such type of 
any other quasi-Wronskian solution in the Moyal deformation. 
Detailed discussion is seen in \cite{Hamanaka8}.
(See also \cite{Paniak}.)

\end{itemize}

\subsection{Reduction of noncommutative anti-self-dual Yang-Mills eq.}

Here we briefly discuss reductions of noncommutative anti-self-dual Yang-Mills equation
into lower-dimensional noncommutative integrable equations such as the 
noncommutative KdV equation. Let us summarize the strategy for
reductions of noncommutative anti-self-dual Yang-Mills equation into lower-dimensions.
Reductions are classified by a choice of gauge group,
a choice of symmetry (such as translational symmetry),
a choice of gauge fixing, and 
a choice of constants of integrations in the process of reductions.
Gauge groups are in general $GL(N)$.
We have to take $U(1)$ part into account in noncommutative case.
A choice of symmetry reduces
noncommutative anti-self-dual Yang-Mills equations to simple forms.
We note that noncommutativity must be eliminated
in the reduced directions
because of compatibility with the symmetry.
Hence within the reduced directions, discussion about the
symmetry is the same as commutative one.

\bigskip

noncommutative anti-self-dual Yang-Mills equations can be represented 
in complex representation as follows (Notation is the same as
the book of Mason-Woodhouse \cite{MaWo}):
\begin{eqnarray}
 F_{wz}&=&\del_{w} A_z -\del_{z} A_w+[A_w,A_z]_\star =0,~
 F_{\tilde{w}\tilde{z}}=\del_{\tilde{w}} A_{\tilde{z}} -\del_{\tilde{z}} A_{\tilde{w}}
 +[A_{\tilde{w}},A_{\tilde{z}}]_\star =0,\nn
 F_{z\tilde{z}}-F_{w\tilde{w}}&=&\del_{z} A_{\tilde{z}} -\del_{\tilde{z}} A_{z}
 +\del_{\tilde{w}} A_{w} -\del_{w} A_{\tilde{w}}
 +[A_z,A_{\tilde{z}}]_\star
 -[A_{w},A_{\tilde{w}}]_\star=0,
\label{asdym}
\end{eqnarray}
where $z,w,{\tilde{z}},{\tilde{w}}$ are linear combinations
of the coordinates of the 4-dimensional spaces $(x^0,x^1,x^2,x^3)$,
and $A_z,A_w,A_{\tilde{z}},A_{\tilde{w}}$ 
denote the gauge fields in the Yang-Mills theory.
This is actually equivalent to the condition of anti-self-duality
of the gauge fields :$F_{\mu\nu}=-*F_{\mu\nu}$ 
where the symbol $*$ is the Hodge dual.

Here, we present non-trivial reductions of noncommutative anti-self-dual Yang-Mills equation with $G=GL(2)$ 
to the noncommutative KdV equation \cite{Hamanaka5}.

First, let us take a dimensional reduction by null translations:
\begin{eqnarray}
 X=\del_w-\del_{\tilde{w}},~Y=\del_{\tilde{z}}.
\end{eqnarray}
and identify space-time coordinates as $t\equiv z,~ x=w+\tilde{w}$,
and put the following non-trivial reduction conditions on the gauge fields:
\begin{eqnarray*}
 \label{kdV}
&A_{\tilde{w}}=\left(\begin{array}{cc}0&0\\u/2&0\end{array}\right), ~
A_{\tilde{z}}=\left(\begin{array}{cc}0&0\\1&0\end{array}\right),~
A_w=\left(\begin{array}{cc}0&-1\\u&0\end{array}\right),~
A_z=\displaystyle\frac{1}{4}
\left(\begin{array}{cc} u^\prime& -2u \\
 u^{\prime\prime}+2u\star u &-u^\prime
\end{array}\right),
\end{eqnarray*}
then we can see Eq. (\ref{asdym}) reduces to the noncommutative KdV equation:
\begin{eqnarray}
 \dot{u}=\displaystyle
\frac{1}{4}u^{\prime\prime\prime}
+\frac{3}{4}\left(u^{\prime}\star u+u\star u^{\prime}\right).
\end{eqnarray}
In this non-trivial way, the noncommutative KdV equation is actually derived.

Many other noncommutative integrable equations are in fact proved to
be derived from noncommutative anti-self-dual Yang-Mills equation by reduction,
which is summarized in \cite{Hamanaka6,Hamanaka7}.
Hence we can make discussion on classification of
lower-dimensional noncommutative integrable equations from
the viewpoint of noncommutative twistor theory.
In particular B\"acklund transformations for noncommutative anti-self-dual Yang-Mills
eqs. are presented \cite{GHN,GHN2} and can be
applied to reveal infinite dimensional symmetry 
of noncommutative anti-self-dual Yang-Mills eqs. and the reduced equations.
Furthermore, this implies that noncommutative integrable equations
such as noncommutative KdV equation can be embedded into a gauge theory
and the Seiberg-Witten map can be applied to them.
The Seiberg-Witten map connects noncommutative description to
commutative one in background of flux, where the degree of
time-derivative is finite. Hence we can discuss
integrability of noncommutative integrable systems via the Seiberg-Witten map. 
These topics are discussed more in detail in \cite{Hamanaka10}.

\section*{Acknowledgements}

The author would like to thank A.~Dimakis, M.~Kato, L.~Mason, 
F.~M\"uller-Hoissen, I.~Strachan, B.~Szablikowski and J.~Wang 
for discussion and comments,
and X. Hu, W. Ma and other organizers for invitation and hospitality
during the World Conference on Nonlinear Analysts 2009 in Orlando
and the International Workshop on Nonlinear and Modern Mathematical
Physics in Beijing. 
He is also grateful to organizers and audiences during the workshops 
YITP-W-10-02 on ``QFT 2010'' for hospitality and discussion.
This work was supported by the Inoue Foundation for Science,
the Daiko Foundation, the Nishina Memorial Foundation,
the Nagoya University Foundation,
and the Showa Public-Reward Foundation.

\begin{appendix}

\section{Brief Introduction to Quasi-determinants}

In the appendix, we make a brief introduction
of quasi-determinants introduced by Gelfand and Retakh
\cite{GeRe} and present a few properties
of them which play important roles in construction
of exact solutions of noncommutative integrable systems.
The detailed discussion is seen in e.g. \cite{GGRW}.

Quasi-determinants are not just a generalization of
usual commutative determinants but rather
related to inverse matrices. From now on,
we suppose existence of all the inverses.

Let $A=(a_{ij})$ be a $N\times N$ matrix and 
$B=(b_{ij})$ be the inverse matrix of $A$,
that is, $A\star B=B\star A =1$.
Here all products of matrix elements are supposed to be
the Moyal-products, though the present discussion hold
for more general situation where the matrix elements
belong to a noncommutative ring.

Quasi-determinants of $A$ are defined formally
as the inverse of the elements of $B=A^{-1}$:
\begin{eqnarray}
 \vert A \vert_{ij}:=b_{ji}^{-1}.
\end{eqnarray}
In the commutative limit, this is reduced to
\begin{eqnarray}
 \vert A \vert_{ij} \stackrel{\theta\rightarrow 0}{\longrightarrow}
  (-1)^{i+j}\frac{\det A}{\det {A}^{ij}},
\label{limit}
\end{eqnarray}
where ${A}^{ij}$ is the matrix obtained from $A$
deleting the $i$-th row and the $j$-th column.

We can write down more explicit form of quasi-determinants.
In order to see it, let us recall the following formula
for a block-decomposed square matrix:
\begin{eqnarray*}
 \left(
 \begin{array}{cc}
  A&B \\C&D
 \end{array}
 \right)^{-1}
=\left(\begin{array}{cc}
 (A-B\star D^{-1}\star C)^{-1}
 &-A^{-1}\star B\star (D-C\star A^{-1} \star B )^{-1}\\
 -(D-C\star A^{-1}\star B)^{-1}\star C\star A^{-1}
&(D-C\star A^{-1}\star B)^{-1}
\end{array}\right),
\end{eqnarray*}
where $A$ and $D$ are square matrices.
We note that any matrix can be decomposed
as a $2\times 2$ matrix by block decomposition
where one of the diagonal parts is $1 \times 1$.
Then the above formula can be applied to the decomposed
$2\times 2$ matrix and an element of the inverse matrix is obtained.
Hence quasi-determinants can be also given iteratively by:
\begin{eqnarray}
 \vert A \vert_{ij}=a_{ij}-\sum_{i^\prime (\neq i), j^\prime (\neq j)}
  a_{ii^\prime} \star (({A}^{ij})^{-1})_{i^\prime j^\prime} \star
  a_{j^\prime
  j}
 =a_{ij}-\sum_{i^\prime (\neq i), j^\prime (\neq j)}
  a_{ii^\prime} \star (\vert {A}^{ij}\vert_{j^\prime i^\prime })^{-1}
  \star a_{j^\prime j}.\nonumber
\end{eqnarray}

It is sometimes convenient to represent the quasi-determinant
as follows:
\begin{eqnarray}
 \vert A\vert_{ij}=
  \begin{array}{|ccccc|}
   a_{11}&\cdots &a_{1j} & \cdots& a_{1n}\\
   \vdots & & \vdots & & \vdots\\
   a_{i1}&~ & {\fbox{$a_{ij}$}}& ~& a_{in}\\
   \vdots & & \vdots & & \vdots\\
   a_{n1}& \cdots & a_{nj}&\cdots & a_{nn}
  \end{array}~.
\end{eqnarray}

Examples of quasi-determinants are,
for a $1\times 1$ matrix $A=a$
 \begin{eqnarray*}
  \vert A \vert  = a,
 \end{eqnarray*}
and 
for a $2\times 2$ matrix $A=(a_{ij})$
 \begin{eqnarray*}
  \vert A \vert_{11}=
   \begin{array}{|cc|}
   \fbox{$a_{11}$} &a_{12} \\a_{21}&a_{22}
   \end{array}
 =a_{11}-a_{12}\star a_{22}^{-1}\star a_{21},~~~
  \vert A \vert_{12}=
   \begin{array}{|cc|}
   a_{11} & \fbox{$a_{12}$} \\a_{21}&a_{22}
   \end{array}
 =a_{12}-a_{11}\star a_{21}^{-1}\star a_{22},\nonumber\\
  \vert A \vert_{21}=
   \begin{array}{|cc|}
   a_{11} &a_{12} \\ \fbox{$a_{21}$}&a_{22}
   \end{array}
 =a_{21}-a_{22}\star a_{12}^{-1}\star a_{11},~~~
  \vert A \vert_{22}=
   \begin{array}{|cc|}
   a_{11} & a_{12} \\a_{21}&\fbox{$a_{22}$}
   \end{array}
 =a_{22}-a_{21}\star a_{11}^{-1}\star a_{12}, 
 \end{eqnarray*}
 and for a $3\times 3$ matrix $A=(a_{ij})$
  \begin{eqnarray*}
  \vert A \vert_{11}
   &=&
   \begin{array}{|ccc|}
   \fbox{$a_{11}$} &a_{12} &a_{13}\\ a_{21}&a_{22}&a_{23}\\a_{31}&a_{32}&a_{33}
   \end{array}
=a_{11}-(a_{12}, a_{13})\star \left(
\begin{array}{cc}a_{22} & a_{23} \\a_{32}&a_{33}\end{array}\right)^{-1}
\star \left(
\begin{array}{c}a_{21} \\a_{31}\end{array}
\right)
\nonumber\\
  &=&a_{11}-a_{12}\star  \begin{array}{|cc|}
                   \fbox{$a_{22}$} & a_{23} \\a_{32}&a_{33}
                   \end{array}^{-1}  \star a_{21}
           -a_{12}\star \begin{array}{|cc|}
                   a_{22} & a_{23} \\\fbox{$a_{32}$}&a_{33}
                   \end{array}^{-1} \star a_{31}      \nonumber\\
&&~~~~    -a_{13}\star \begin{array}{|cc|}
                   a_{22} & a_{23} \\\fbox{$a_{32}$}&a_{33}
                                         \end{array}^{-1}\star  a_{21}
           -a_{13}\star \begin{array}{|cc|}
                   a_{22} & a_{23} \\a_{32}&\fbox{$a_{33}$}
                   \end{array}^{-1} \star a_{31},
 \end{eqnarray*}
and so on.

\end{appendix}

\end{document}